\documentclass[11pt,a4paper,twoside,groupcitations]{article}
\usepackage[T1]{fontenc}
\usepackage[ansinew]{inputenc}
\usepackage[english]{babel}
\usepackage{amsfonts}
\usepackage{amsmath}
\usepackage{bm}
\usepackage{array}
\usepackage{amsthm}
\usepackage{wasysym}
\usepackage{amssymb}
\usepackage{graphicx}
\usepackage{subfigure}
\usepackage{bigints}
\usepackage{braket}
\usepackage{eucal}
\usepackage{float}
\usepackage{verbatim}
\usepackage[table]{xcolor}
\usepackage{caption}
\usepackage{textcomp}
\usepackage{hyperref}
\usepackage{color}
\usepackage{mathtools}
\usepackage{commath}
\usepackage{bigints}
\usepackage[toc,page]{appendix}
\usepackage{csquotes}
\usepackage{multicol}
\usepackage{tikz}
\usetikzlibrary{positioning,arrows}
\usetikzlibrary{decorations.pathmorphing}
\usetikzlibrary{decorations.markings}
\usetikzlibrary{calc,decorations.markings}
\usetikzlibrary{arrows,shapes}
\usetikzlibrary{matrix,arrows}
\usepackage{pgfplots}
\usepackage{xparse}
\usepackage[export]{adjustbox}
\definecolor{jade}{HTML}{00A86B}
\newcommand{\be}{\begin{eqnarray}}
\newcommand{\ee}{\end{eqnarray}}

\renewcommand{\d}{\mbox{${\rm d}$}} 
\newcommand{\lp}{\ell_{\rm p}}
\newcommand{\mpl}{m_{\rm p}}
\newcommand{\gn}{G_{\rm N}}

\newcommand{\Rh}{R_{\rm H}}

\graphicspath{{Figures/}}
\begin{document}
\title{\bf Quantum dust cores of black holes and their quasi-normal modes}
\author{T.~Bambagiotti$^{bd}$\thanks{E-mail:
tommaso.bambagiotti2@unibo.it},
L.~Gallerani$^{ac}$\thanks{E-mail: luca.gallerani6@unibo.it},
A.~Mentrelli$^{acd}$\thanks{E-mail: andrea.mentrelli@unibo.it},
A.~Giusti$^{bcd}$\thanks{E-mail: andrea.giusti9@unibo.it},
and
R.~Casadio$^{bcd}$\thanks{E-mail: casadio@bo.infn.it}
\\
\\
$^a${\em Department of Mathematics, University of Bologna}
\\
{\em Piazza di Porta San Donato 5, 40126 Bologna, Italy}
\\
\\
$^b${\em Department of Physics and Astronomy "A. Righi", University of Bologna}
\\
{\em via Irnerio~46, 40126 Bologna, Italy}
\\
\\
$^c${\em Alma Mater Research Center on Applied Mathematics (AM$^2$)}
\\
{\em Via Saragozza 8, 40123 Bologna, Italy}
\\
\\
$^d${\em I.N.F.N., Sezione di Bologna, I.S.~FLAG}
\\
{\em viale Berti~Pichat~6/2, 40127 Bologna, Italy}
}
\maketitle
\begin{abstract}
The quantum description of a gravitationally collapsed ball of dust proposed in
Ref.~\cite{Casadio:2023ymt} is characterised by a linear effective Misner-Sharp-Hernandez
mass function describing a matter core hidden by the event horizon.
After reviewing the original model and some of its refinements, we investigate the quasi-normal mode
spectrum of the resulting spacetime and compare it with the Schwarzschild case.
Computations are performed within the WKB approximation, based on the Pad\'e
approximants up to thirteenth order.
Our analysis shows that deviations from the Schwarzschild spectrum are sensitive to the
quantum nature of the core surface.
\end{abstract}
%
%
%
%
%
%
\section{Introduction}
\label{sec1}
\setcounter{equation}{0}
Black holes are some of the most challenging physical prediction of General Relativity (GR).
Traditionally, they are described by stationary vacuum solutions of the Einstein equations,
which are characterised by the presence of an event horizon and by their geodesic incompleteness.
Spacetime singularities mark the breakdown of the classical theory of gravity, and there is a growing
consensus that they should not show up in a complete, but still unknown, theory of quantum
gravity~\cite{Hawking:1973uf}.
Moreover, their trivial vacuum structure is likely altered if one takes into account
the quantum nature of matter that collapses and forms astrophysical black holes.
\par
Fundamental constituents of matter in the standard model of particle physics are represented
by quantum excitations of fields that behave like localised objects (particles) in suitable regimes
(when particle number is conserved), and any realistic modelling of black holes should
somehow take into account these features.
In this context, a quantum model was studied in Ref.~\cite{Casadio:2023ymt} in which a black hole forms
from the gravitational collapse of an isotropic distribution of dust,
partitioned into nested layers of dust particles moving in the Schwarzschild spacetime
\be
\label{Schw}
\d s^2
=
-\left(1-\frac{2\,\gn\, m}{r} \right)\d t^2
+ \left(1-\frac{2\,\gn\, m}{r} \right)^{-1}\d r^2
+r^2\,\d \Omega
\ ,
\ee
where $m=m(r)$ is the Misner-Sharp-Hernandez (MSH) fraction of the Arnowitt-Deser-Misner (ADM)
mass inside a sphere of radius $r=r(\tau)$ and
$\d\Omega^2=\d \theta^2+\sin^2{\theta}\,\d\varphi^2$.~\footnote{We shall always use units with $c = 1$ and often
write the Planck constant $\hbar = \lp \, \mpl $ and the Newton constant $\gn = \lp/\mpl$,
where $\lp$ and $\mpl$ are the Planck length and mass, respectively.}
Each dust particle carries the same proper mass $\mu$, while the $i^\mathrm{th}$ layer has a
MSH mass $\mu_i$ for $i=0,\ldots,N$, where $i=0$ refers to the innermost core, and $i=N$ labels the
outermost shell.
The cumulative mass $M_i$ inside the $i^{\rm th}$ layer is then given by
\be
M_i
=
\sum_{j=0}^{i-1}\mu_j
\ ,
\ee
with $M_{1}=\mu_0$ and the total ADM mass $M=M_{N+1}$.
During the collapse, each dust particle falls freely along radial time-like geodesics, whose equation
for the shell $r=R_i(t)$ reads
\be
\label{H}
H_i
\equiv 
\frac{P^2_i}{2\,\mu}
-\frac{\gn\, \mu\, M_i}{R_i}
=
\frac{\mu}{2}\left(\frac{E^2_i}{\mu^2}-1 \right)
=
\mathcal{E}_i,
\ee
where $H_i$ denotes the Hamiltonian of the system, $P_i=\mu \, \d R_i/\d  \tau$ is the radial momentum
conjugated to $R_i(\tau)$ and $E_i$ the conserved momentum conjugated to $t = t_i(\tau)$.
\par
\begin{figure}[t]
\centering
\begin{minipage}{.5\textwidth}
\centering
\includegraphics[width=\linewidth, height=0.25\textheight]{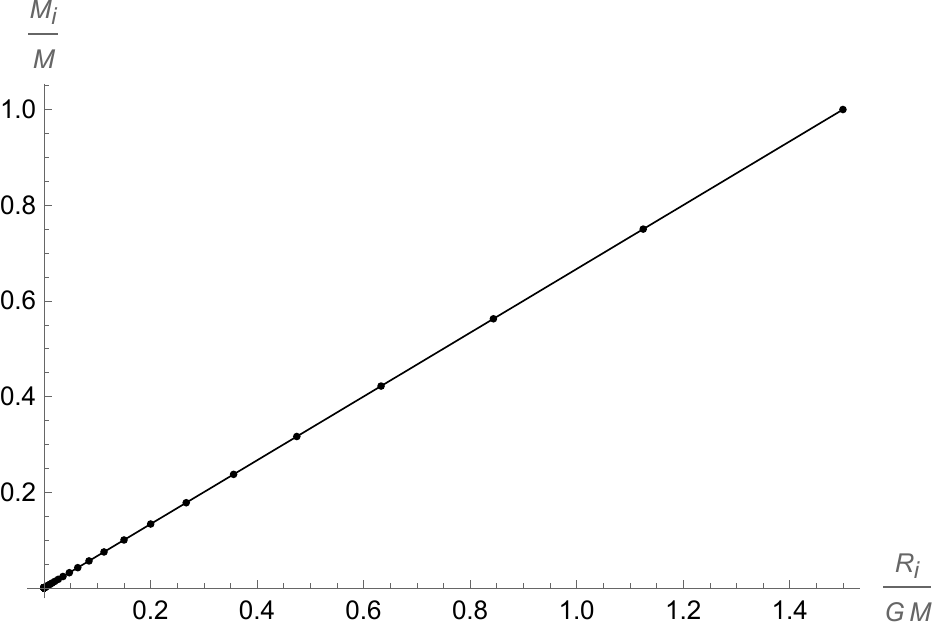}
\end{minipage}%
\begin{minipage}{.5\textwidth}
\centering
\includegraphics[width=\linewidth, height=0.25\textheight]{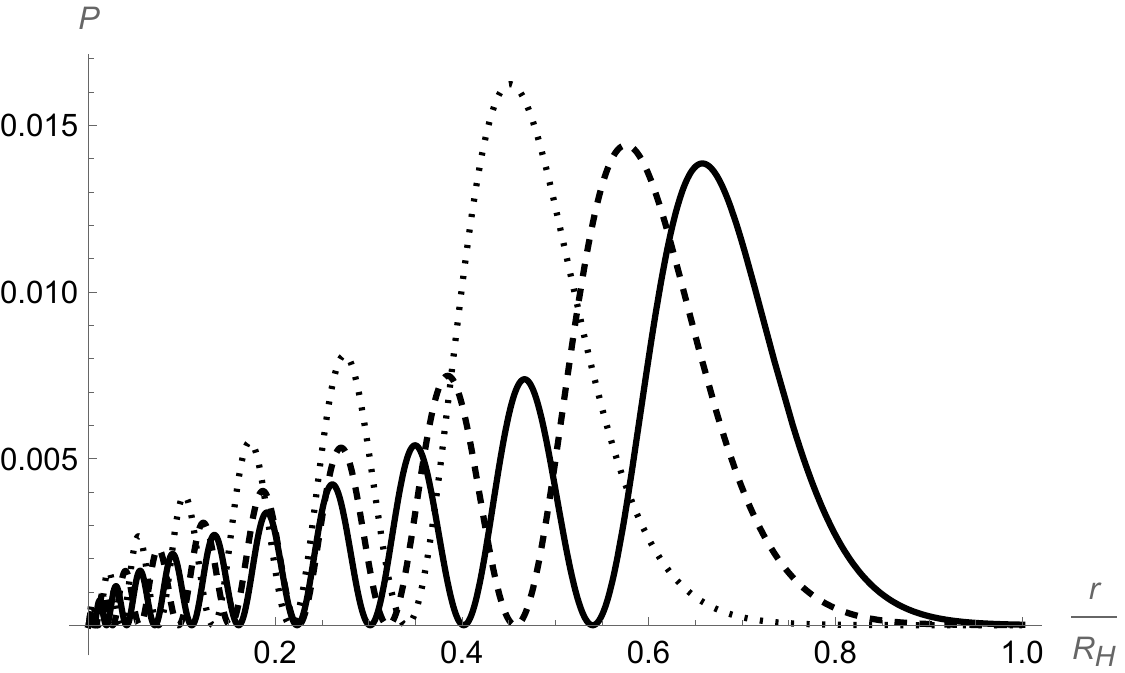}
\end{minipage}
\caption{Left panel: discrete mass function $M_i$ (dots) for $N=100$ layers
and its continuous approximation~\eqref{mlinear} (solid line).
Right panel: probability densities~\eqref{P} for $N=3$, $\mu = {\mpl}/{10}$, $M={(440/3)\,\mu}$.}
\label{N=3 & linear fit}
\end{figure}

The differences between our description of a ball of dust particles and other approaches to the
quantisation of a dust ball are discussed in details in Ref.~\cite{Casadio:2023ymt}.
We just remark that our procedure amounts to quantising the trajectories of individual dust particles like is done
for the electron in the quantum mechanics of the hydrogen atom, whereas canonical approaches
like that in Ref.~\cite{Vaz:2011zz} quantise collective degrees of freedom, such as the radius of the ball or of shells.
The relativistic action for geodesics of dust particles is their proper time (times the mass $\mu$),
which yields the mass-shell condition in the form of the Hamiltonian constraint~\eqref{H}~\cite{kiefer}.
The canonical quantisation prescription therefore leads to a time-independent Schr\"odinger equation,
\be
\hat H_i\,
\psi_{n_i}
=
\mathcal E_i\,\psi_{n_i}
\ ,
\ee
whose solutions are given by the Hamiltonian eigenfunctions $\psi_{n_{i}}$ corresponding to the eigenvalues
\be
\mathcal{E}_i
=
\frac{\mu^3\, M^2_i}{2 \,\mpl^4\, n^2_i}
\ .
\label{eigenvalues}
\ee
We also recall that the measure in this Hilbert space is chosen to ensure that $\hat H_i$ are Hermitian
(precisely like for the hydrogen atom)~\cite{Casadio:2023ymt}.
Since $E_i^2\ge 0$, Eq.~\eqref{H} sets the bound $n_{i}\geq N_{i}$, where~\footnote{See also Ref.~\cite{Casadio:2021cbv}.} 
\be
N_i
=
\frac{\mu \,  M_{i}}{\mpl^2}
\label{N_i}
\ee
represents the allowed ground state of each particle in the $i^{\rm{th}}$ layer.
The corresponding wave-function is given by
\be
\label{psi}
\psi_{N_i}(R_i)
=
\sqrt{\frac{\mu \, \mpl}{\pi\, \lp^3 \, M^{2}_{i} }}\,
\exp{\left(-\frac{\mu \, R_i}{ \mpl \, \lp} \right)}\,
L^1_{N_i-1}\!\left(\frac{2 \,\mu \, R_i}{ \mpl \,\lp} \right)
\ ,
\ee
with $L^{1}_{n-1}$ the generalised Laguerre polynomials and $n=1,2,\ldots$.
Finally, the wave-functions $\eqref{psi}$ are normalised in the scalar product which makes $\hat{H}_i$ Hermitian:
\be
\braket{n_i|n'_i}
=
4\,\pi \int_0^{\infty}
R^2_i \, \psi^{*}_{n_i}(R_i)\psi_{n_{i}'}(R_i)\, \d R_i
=
\delta_{n_in'_i}
\ .
\ee
For this ground state, one can compute both the expectation value
\be
\bar{R}_i 
=
\braket{N_i|\hat{R}_i|N_i}
=
\frac{3}{2}\,\gn\, M_i
\label{Rmean}
\ee
and its uncertainty
\be
\frac{\Delta{R}_i}{\bar{R}_i}
= 
\frac{\sqrt{\braket{N_i|\hat{R}^2_i|N_i}-\bar{R}^2_i}}{\bar{R}_i}
=
\frac{\sqrt{N_i^2+2}}{3N_i}
\simeq
\frac{1}{3}
\ ,
\label{DR}
\ee
where we assumed $N_i \gg 1$.
The finest layering compatible with the quantum uncertainty~\eqref{DR} is given by
$\bar{R}_{i+1} \simeq \bar{R}_{i} + \Delta {R}_i$.
The discrete MSH mass is therefore distributed as $M_{i+1}\simeq 4\,M_i/3$,
which means that $M_i$ grows linearly with respect to the radius $R_i$
(regardless of the total number of layers $N$).
This result motivates the introduction of an effective energy density
\be
\label{rho}
\rho 
\simeq
\frac{\mpl}{6\, \pi \,\lp \,r^2}
\ ,
\ee
which yields a linear effective MSH mass
\be
\label{mlinear}
m(r)
\simeq
\frac{2 \,\mpl \, r}{3 \, \lp}
\ .
\ee
This mass function weakens the central singularity of the Schwarzschild solution and matches the total ADM mass
at the surface of the outermost layer, as shown in the left panel of Fig.~\ref{N=3 & linear fit}.
Eq.~\eqref{mlinear} thus provides an effective description of the geometry of a collapsed core of
radius $R_{\rm s} \simeq \bar{R}_N + \Delta {R}_N \simeq 3\,\gn\,M/2$, shorter than, but still comparable
to, the gravitational radius $\Rh = 2\, \gn\, M$ (more details about the effective energy-momentum tensor of the ground state and
interior geometry can be found in Ref.~\cite{Casadio:2023ymt}).
\par
This brief summary of the model from Ref.~\cite{Casadio:2023ymt} illustrates the derivation of its linear
mass distribution~\eqref{mlinear}, which regularises the singularity and represents an interesting playground
for the calculation of quasi-normal modes (QNMs).
We recall that black holes are characterised by quasi-normal frequencies of
oscillation~\cite{Berti:2009kk, Nollert:1999ji, Kokkotas:1999bd} and footprints of the collapsed dust distribution
could then be observed in deviations from the characteristic spectrum of a classical Schwarzschild black hole. 
In the analysis of linear perturbations of a static and spherically symmetric gravitational background,
QNMs are solutions of the form $e^{-i\,\omega\,t}\,\xi_{\omega}(r)$ which satisfy the boundary conditions
\begin{align}
\label{eq:qnm_boundary}
\xi_{\omega}(r_{*})
\sim
\begin{cases}
\exp(-i\,\omega\, r_{*})
\ ,
\qquad 
&{\rm for}
\quad
r_{*} \to -\infty
\\[5pt]
\exp(+i\,\omega\, r_{*})
\ ,
\qquad 
&{\rm for}
\quad
r_{*} \to +\infty
\ ,
\end{cases}
\end{align}
where $r_*$ is the tortoise radial coordinate.
The outer horizon forbids the system to be time-symmetric,
hence the boundary problem to be Hermitian, and this results in complex frequency at infinity,
\be
\omega = \omega_R +i \, \omega_I
\ .
\ee
Since the boundary problem~\eqref{eq:qnm_boundary} admits only discrete values with imaginary
part $\omega_I<0$, quasi-normal modes describe damped oscillations with a decay timescale set by $\omega_I^{-1}$.
\par
In Section~\ref{sec2}, we will show that the linear mass profile can be improved by taking into account
the effective quantum description of matter in the outermost layer or among all layers of the collapsed
core~\cite{Gallerani:2025wjc};
QNMs are then computed for each different description of the core in Section~\ref{sec3},
and conclusions are drawn in Section~\ref{sec4}.
\section{Mass distributions}
\label{sec2}
\setcounter{equation}{0}
The linear effective mass function~\eqref{mlinear} matches continuously with the outer Schwarzschild
geometry of constant ADM mass $M$ (for $r\gtrsim R_{\rm s}$).
However, its first derivative is discontinuous at the core surface, with $m'(r)>0$ for $r \to R_{\rm s}^{-}$
[and $m'(r)=0$ for $r > R_{\rm s}$], which arguably conflicts with the quantum core being in (final)
equilibrium in the ground state.
We also remark that the outermost layer with (approximate) thickness $\Delta {R}_{N} \simeq R_{\rm s}/4$
contains one  fourth of the total mass $M$,
\be
\mu_N
=
M_{N+1} - M_{N}
\simeq
\frac{M}{4}
\ .
\ee
A better approximation for the effective geometry near the core surface can be obtained by considering
more carefully the fuzziness of quantum layers, which is effectively described by the thickness $\Delta {R}_{i}$
in Eq.~\eqref{DR}.
\par
A more accurate description of the mass distribution inside each layer can be obtained from the probability
density to localise a particle of the $i^{\rm th}$ layer:
\be
\label{P}
\mathcal{P}_i
=
4\,\pi \, r^2 \, |\psi_{n_i}(r)|^2
\ .
\ee
The corresponding effective mass distribution is then given by $\eta_i = \mu_i\,\mathcal{P}_i$,
which also shows a non-zero probability to find a particle in a different layer $j \neq i$.
In particular, the mass distribution in the outermost layer 
\be
\eta_N = \mu_N\,\mathcal{P}_N
\label{etaN}
\ee
does not depend on the number of layers $N$, and is again expected to match smoothly with the
outer Schwarzschild geometry of total ADM mass $M$.
This is only possible if the behaviour of $\eta_N$ deviates from the linear form in~\eqref{mlinear},
which also follows from the wave-functions~\eqref{psi}.
\par
Finally, we note that the same argument based on the wave-functions~\eqref{psi} implies
that dust particles can be located at positions $r>\Rh$ with finite (albeit typically very small) probability..
\subsection{Quantum mass refinement} 
\label{sec2.1}
The wave-function~\eqref{psi} of each layer shows an oscillatory damped pattern that suggests
an overlap among (at least) the nearest neighbours.
This effect turns the linear mass profile into a (slightly) parabolic mass distribution inside the
ball (all the details can be found in Ref.~\cite{Gallerani:2025wjc}).
In fact, particles from the $i^{\rm{th}}$ layer have non-vanishing probability to be located inside
the $j^{\rm{th}}$ layer given by Eq.~\eqref{P} (see right panel of Fig.~\ref{N=3 & linear fit}).
\par
\begin{figure}[t]
\centering
\begin{minipage}{.5\textwidth}
\centering
\includegraphics[width=\linewidth, height=0.25\textheight]{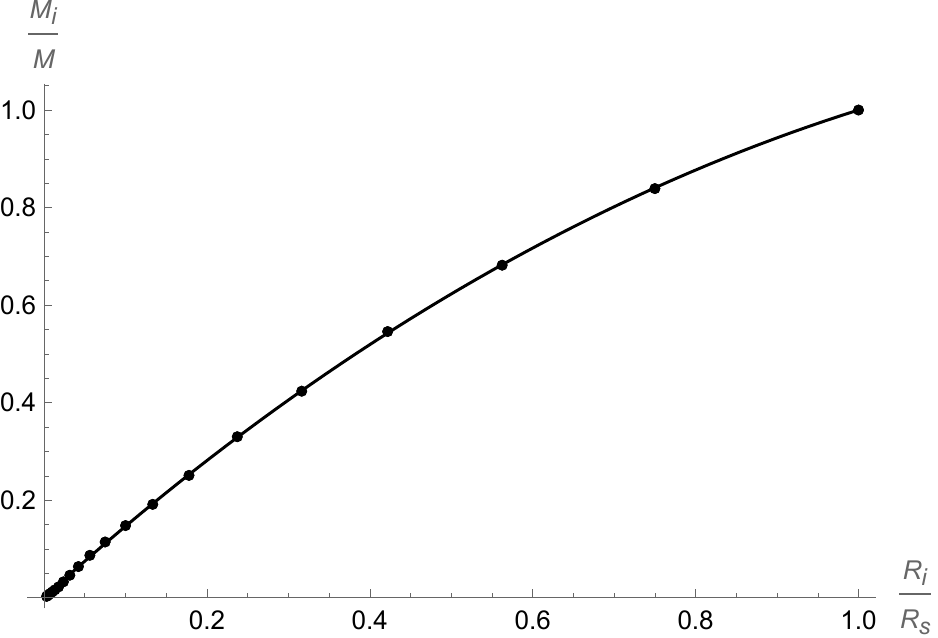}
\end{minipage}%
\begin{minipage}{.5\textwidth}
\centering
\includegraphics[width=\linewidth, height=0.25\textheight]{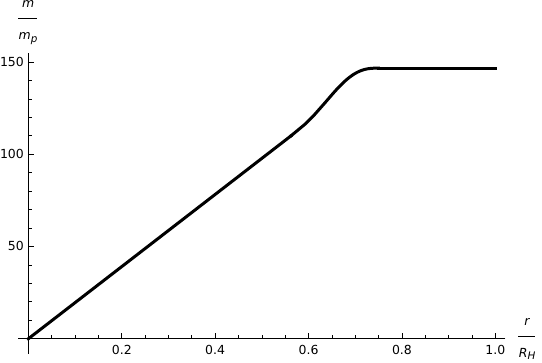}
\end{minipage}
\caption{Left panel: 
quantum corrected MSH mass function for $N=20$ with $M=3000 \, \mpl$ and $\mu=\mpl/10$.
Right panel: 
interpolating MSH mass function for a black hole of total ADM mass $M \simeq 150 \,\mpl$,
$\mu = \mpl/10$ and $\Rh = 300 \,\lp$.}
\label{mass profiles}
\end{figure}

The mass elements which take into account this quantum effect are defined as
\be
\Delta \mu_{i,j}
=
\mu_i \int_{R_j}^{R_{j+1}}\mathcal{P}_i(r)\,\d r
\ ,
\quad \quad
i,j=0,\ldots,N
\ .
\ee
The sum of all these contributions inside the $j^{\mathrm{th}}$ layer gives the refined mass of
that layer indicated as $\Delta \mu_j$.
The total mass distribution $\mathcal{M}_i$ is then obtained by cumulative summing the mass of each layer:
\be
\label{Mnew}
\mathcal{M}_i
=
\sum_{j=0}^{i}
\Delta \mu_j
\ ,
\quad \quad
i=0,\ldots,N
\ .
\ee
The best fit of the curve obtained from the numerical calculation of $\mathcal{M}_i$ shows the parabolic profile
\be
\label{Mparabolic}
m
=
M
\left(a\,x + b \, x^c\right)
\ ,
\ee
with $x=r/R_H$ and $a$, $b$, $c$ fitting coefficients~\cite{Gallerani:2025wjc}.~\footnote{The
profile does not significantly depend on the number $N$ of layers since the largest faction of the total
mass is located in the outer layers.}
This refined distribution preserves the integrability of the metric in the origin and does not give rise
to any inner horizons.
Values of $\mathcal{M}_i$ and the parabolic profile are shown in the left panel of Fig.~\ref{mass profiles}.
\par
It is important to recall that the ground state for the dust core that we consider requires a large
number of dust particles in each layer, which means that $M_i\gg \mpl\gtrsim \mu$.
For a realistic astrophysical black hole, with $M$ at least of the order of the solar mass and $\mu$
of the order the proton mass, the Laguerre polynomials in the wave-functions~\eqref{psi} are of
order $n\gtrsim 10^{18}$, which is numerically intractable.
For this reason, in the following, we shall perform all numerical calculations that involve
Eq.~\eqref{psi} only for values of $M_i$ and $\mu$ that are tractable, albeit far from
realistic.
\subsection{Boundary layer}
\label{sec2.2}
The above improved description of the effective mass function allows us to introduce equally 
improved descriptions of the outermost layer of dust based on the wave-function~\eqref{psi}
for $i=N$.
\par
First of all, for the linear mass distribution~\eqref{mlinear}, we can consider a smoothed MSH mass
for the outermost layer given by
\be
m_{\rm lin}(r\geq R_N)
&\!\!=\!\!&
M_{N}
+ 
\int_{\bar R_N}^{r} \eta_N(x)\,\d x
\nonumber
\\
&\!\!=\!\!&
M_{N}
+ 
4\, \pi \,\mu_N
\int_{\bar R_N}^{r} x^2\, |\psi_N(x)|^2\,\d x
\ ,
\label{m_lin}
\ee
with $M_N=3\,M/4$ and $\mu_N=M/4$.
This function accounts for the linear mass profile up to the inner border $r=\bar R_N$ of the outermost layer,
but is also sensitive to the quantum nature of the dust particles which can be localised outside $R_{\rm s}$,
an effect particularly relevant in this (most massive) layer.
\par
Similarly, for the parabolic refinement~\eqref{Mparabolic} reviewed in Section~\ref{sec2.1}, we consider
a MSH function for the outermost layer given by
\be
m_{\rm par}(r\geq R_N )
=
\mathcal{M}_N
+ 
4\, \pi \,\mu_N
\int_{\bar R_N}^{r} x^2\, |\psi_N(x)|^2\,\d x
\ ,
\label{m_par}
\ee
where $\mathcal{M}_N$ and the mass of the outermost layer $\mu_N$ are now obtained using the
numerical mass profile~\eqref{Mnew}. 
\par
In both cases above, the effective MSH mass approaches $M$ more rapidly (and smoothly)
than the simple linear function~\eqref{mlinear} in the transitional region of width $\Delta {R}_{N}$.
We can also consider a mass function $m=m(r)$ which interpolates smoothly between the linear 
form~\eqref{mlinear} for $r \lesssim r_0\equiv\bar{R}_{N}$ and the outer Schwarzschild vacuum for
$r>r_1\equiv \bar{R}_{N}+\Delta {R}_{N}=R_{\rm s}$.
Clearly, $r_0$ and $r_1$ parametrise the boundaries of the outermost layer
and we can write the interpolating mass function as
\begin{equation}
\label{eq:massfunctioninterpolated}
m_{\rm int}(r)
=
\begin{cases}
\alpha \, r 
\ ,
\qquad 
&{\rm for}
\quad
r \leq r_0
\\ 
B(r)
\ ,
\qquad 
&{\rm for}
\quad
r_0 \leq r \leq r_1
\\
M
\ ,
\qquad 
&{\rm for}
\quad
r \geq r_1
\ ,
\end{cases}
\end{equation}
where $\alpha$ is a constant and $B(r)$ a function to be determined.
The right panel of Fig.~\ref{mass profiles} shows the interpolating mass function~\eqref{eq:massfunctioninterpolated},
whose analytical expression can be found in Appendix~\ref{appendix-a} (see also Ref.~\cite{TBms}). 
As required, the interpolating mass function~\eqref{eq:massfunctioninterpolated} matches the outer Schwarzschild 
geometry, with continuous first and second derivatives at $r=R_{\rm s}$.
This implies that quantum effects at the outermost layer are neglected in this approximation
and the corresponding outer geometry is exactly given by the Schwarzschild vacuum.
In this case, no effect can be expected on linear perturbations for $r>\Rh$, including the QNMs that we are going to study
in the next Section.
\section{Quasi-normal modes}
\label{sec3}
\setcounter{equation}{0}
Let us briefly recall the theoretical framework of black hole QNMs.
Their governing equation can be solved with various numerical methods, extensively reviewed
in Ref.~\cite{Konoplya:2011qq}, but we shall here only employ the WKB approach.
\par
For a generic static and spherical symmetric spacetime, the metric reads
\be
\label{sss}
\d s^2
=
-f(r)\,
\d t^2
+
h(r)
\,
\d r^2
+
\sigma^2(r)
\,
\d\Omega^2
\ .
\ee
Scalar (massless) perturbations are described by solutions of the Klein-Gordon equation
\be
\label{KG}
\Box\Phi=\frac{1}{\sqrt{-g}}
\,
\partial_{\mu}
\,
[\sqrt{-g}
\, 
g^{\mu \nu}
\,
\partial_{\nu}]
\,
\Phi= 0
\ ,
\ee
and spherical symmetry and time independence of the metric allow for the factorisation
\be
\Phi
=
e^{-i\,\omega\, t} \, \phi_{\omega \ell m} (r) \, Y_l^m
\ ,
\ee
where $Y_{\ell}^m=Y_{\ell}^m(\theta,\phi)$ are spherical harmonics and $\phi_{\omega \ell m}$
is the radial function with $0\le l\le n$ and $-l\le m\le l$.
Eq.~\ref{KG} then yields
\be
\omega^2\,\phi_{\omega \ell m}
+
\frac{f}{h}\left[\phi''_{\omega \ell m}+\left(\frac{2\,\sigma'}{\sigma}+\frac{f'}{2\,f}-\frac{h'}{2\,h}\right)\phi'_{\omega \ell m}\right]
-
\frac{\ell\,(\ell+1)}{\sigma^2}\,f\,\phi_{\omega \ell m}
=
0.
\ee
It is convenient to perform the substitution
\be
\phi_{\omega \ell m}(r)
=
\left(\frac{h}{f}\right)^{1/4}\frac{\xi}{\sigma}
\ ,
\ee
with $\xi=\xi_{\omega \ell m}(r)$, that leads to
\be
\omega^2\,\xi
&\!\!=\!\!&
-
\frac{f}{h}
\left\{
\xi''
+
\frac{1}{4}
\left[
\frac{h''}{h}
-
\frac{f''}{f}
-\frac{4\,\sigma''}{\sigma}
+
\frac{3}{4}\left(\frac{f'}{f}\right)^2
-
\frac{5}{4}\left(\frac{h'}{h}\right)^2
-
\frac{2\,\sigma'\,f'}{\sigma\,f}
+
\frac{2\,\sigma'\,h'}{\sigma\,h}
+
\frac{f'\,h'}{2\,f\,h}
\right]
\xi
\right\}
\nonumber
\\
&&
+
\frac{\ell\,(\ell+1)}{\sigma^2}\,f\, \xi
\ .
\label{KGu}
\ee
This can be greatly simplified by setting $h=f^{-1}$, using areal coordinates $\sigma=r$,
and introducing the tortoise coordinate $\d r_*=\d r/f$.
Eq~\eqref{KGu}, finally becomes
\be
\frac{d^2\xi(r)}{d r^2_*}
+
\left[
\omega^2
-
V(r)
\right]
\xi(r)
=
0
\label{KGuT}
\ee
with
\be
V(r)=\frac{\partial^2_{r_*}r}{r}+\frac{\ell(\ell+1)}{r^2}\,f(r)
\ ,
\ee
where $r=r(r_*)$.
Similar equations hold for vector and tensor perturbations and a master (radial) equation
can be cast in the form 
\be
\frac{d^2\xi_s(r)}{d r^2_*}+Q_s(r)\,\xi_s(r)
=
0
\ ,
\label{KGuS}
\ee
with $Q_s(r)=\omega^2-V_s(r)$ and
\be
V_s(r)
=
f(r)
\left[\frac{\ell(\ell+1)}{r^2}+\frac{f'(r)}{r}(1-s^2)\right]
\ ,
\label{Vs}
\ee
where the label $s$ refers to scalar ($s=0$), vector ($s=1$) and tensor ($s=2$) perturbations.
\par
This definition holds for $s = 0, 1, 2$, in the Schwarzschild space-time.
However, for $s = 2$, the tensor perturbations admit a multipole decomposition yielding an odd-parity
(axial) and even-parity (polar) potential.
The former, also called Regge-Wheeler (RW) potential, coincides with the one obtained from~\eqref{Vs} for $s=2$,
namely
\be
V_2^{(\mathrm{o})}(r)
=
\left(1-\frac{\Rh}{r}\right)\left[\frac{\ell(\ell+1)}{r^2}-\frac{3\,\Rh}{r^2}\right]
\ ,
\label{RW}
\ee
while the latter, known as Zerilli potential, takes the form
\be
V_2^{(\mathrm{e})}(r)
=
\frac{2}{r^3}\left(1-\frac{\Rh}{r}\right)
\frac{
9\,\gn^3\,M^3
+3\,\lambda^2\,\gn\,M\,r^2+\lambda^2(1+\lambda)\, r^3+9\, \gn^2\,M^2 \lambda\,r}
{(3\,\gn\,M+\lambda \, r)^2}
\ ,
\label{Ve}
\ee
with $\lambda\equiv (\ell-1)(\ell+2)/2$.
\par
\begin{figure}[t]
\centering
\includegraphics[width=0.43\textwidth,height=0.3\textwidth]{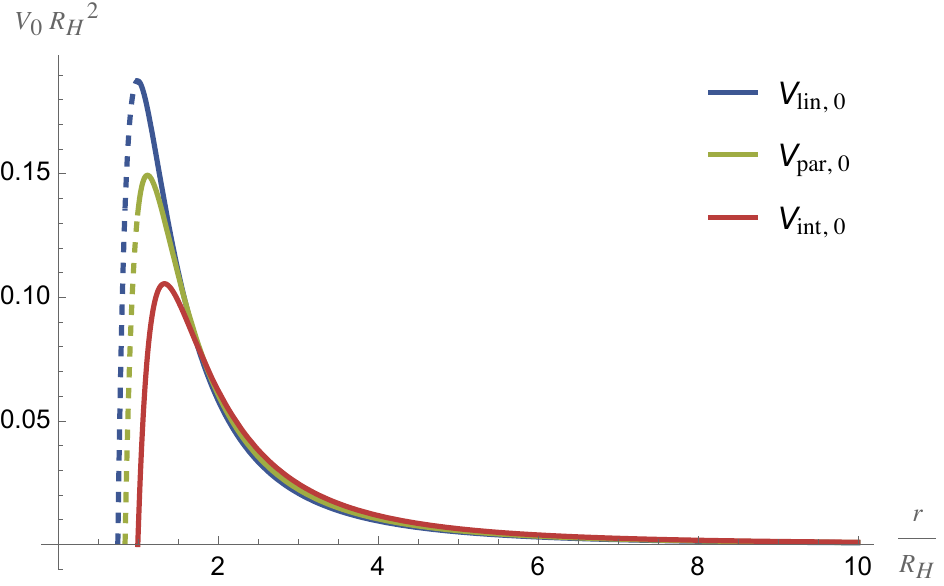}
$\qquad$
\includegraphics[width=0.43\textwidth,height=0.3\textwidth]{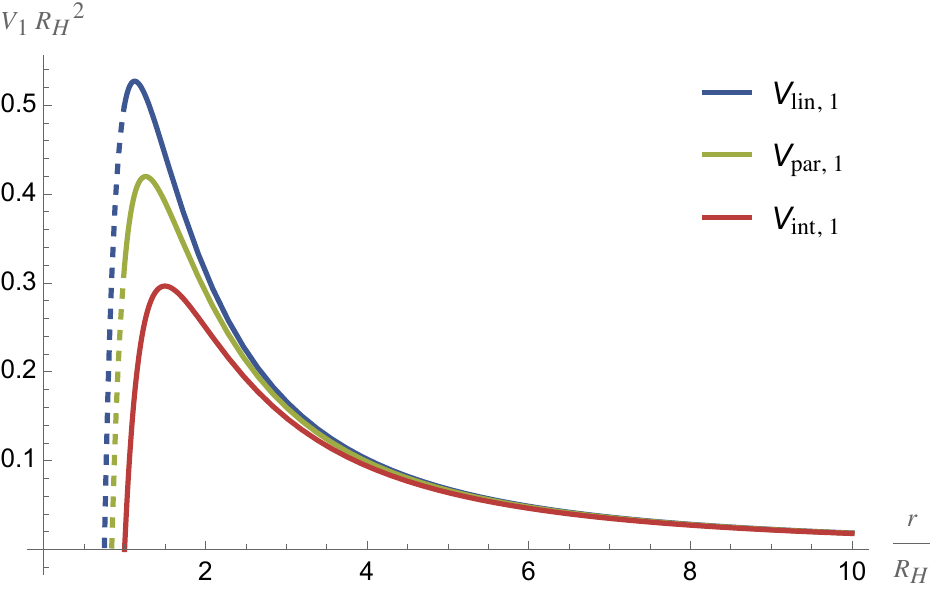}
\\
$\ $
\\
\includegraphics[width=0.43\textwidth,height=0.3\textwidth]{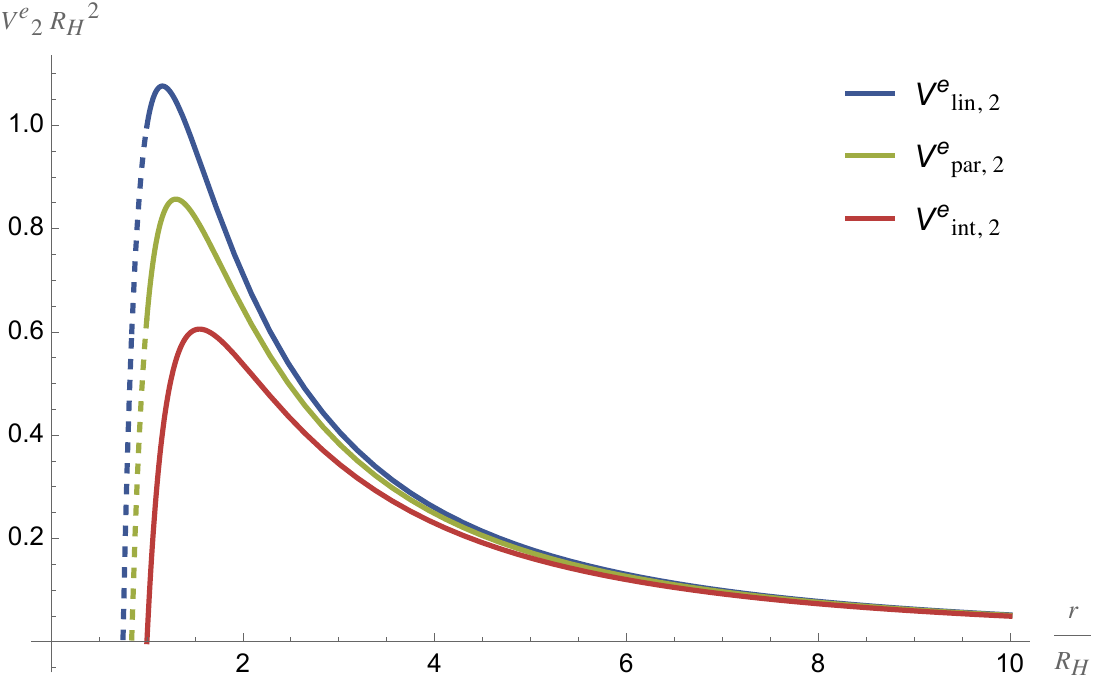}
$\qquad$
\includegraphics[width=0.43\textwidth,height=0.3\textwidth]{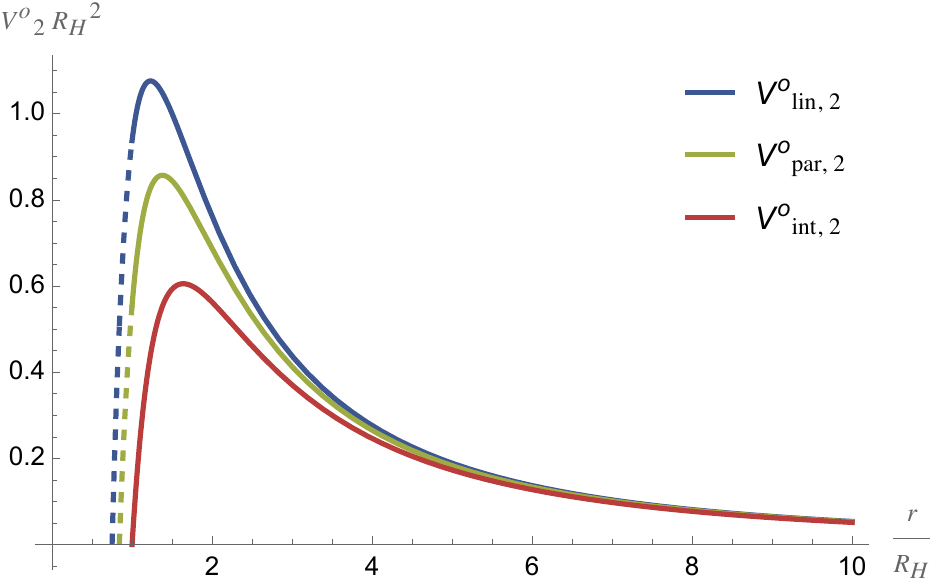}
\caption{Potentials for scalar perturbations $V_0\,\Rh^2$ (top left), vector perturbations $V_1\,\Rh^2$ (top right),
even tensor perturbations $V^{\mathrm{(e)}}_2\, \Rh^2$ (bottom left), and odd tensor perturbations
$V^{\mathrm{(o)}}_2\,\Rh^2$ (bottom right): $V_{\rm lin}$ for the linear mass function~\eqref{m_lin},
$V_{\rm par}$ for the parabolic mass function~\eqref{m_par}, and $V_{\rm int}$ for the interpolating
mass function~\eqref{eq:massfunctioninterpolated}.
All plots with $\mu=M/10=\mpl/10$.}
\label{fig:V}
\end{figure}
\begin{figure}[t]
\centering
\includegraphics[width=0.43\textwidth,height=0.3\textwidth]{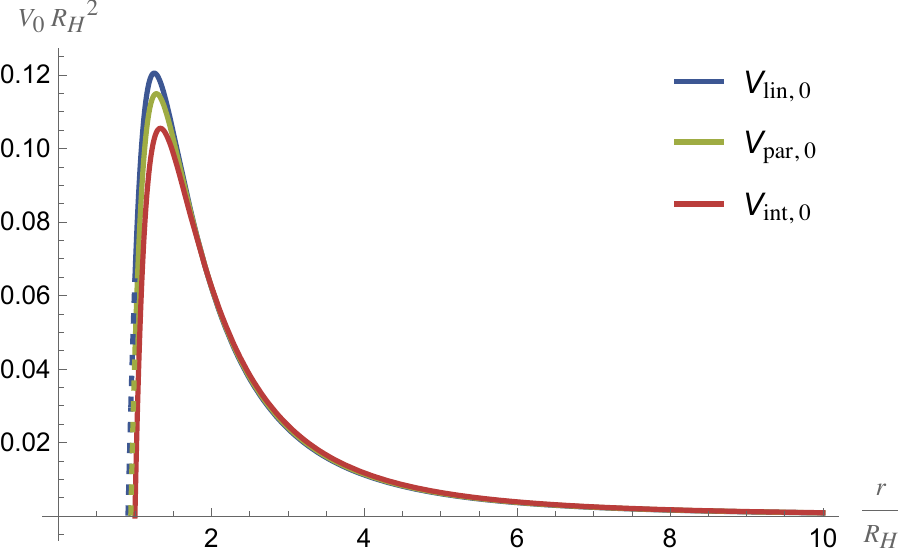}
$\qquad$
\includegraphics[width=0.43\textwidth,height=0.3\textwidth]{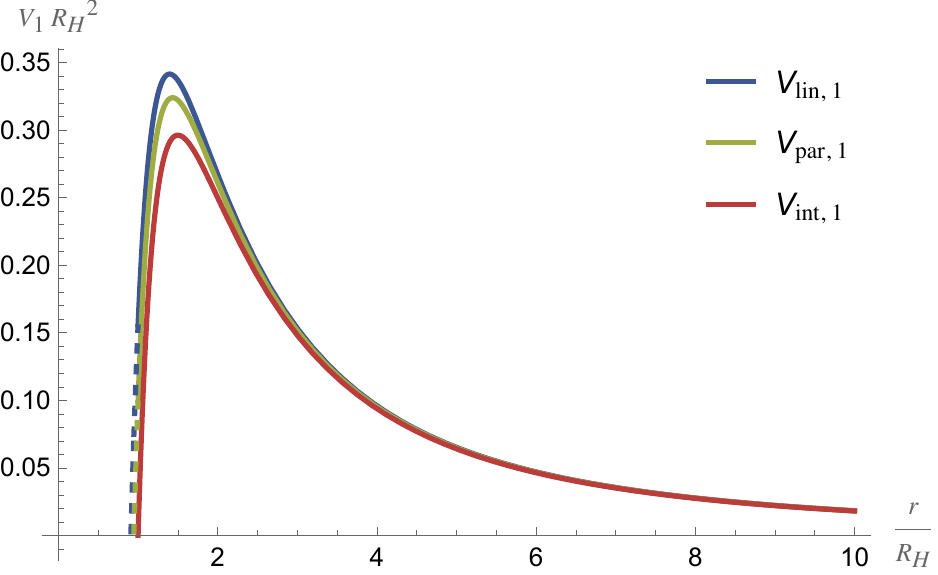}
\\
$\ $
\\
\includegraphics[width=0.43\textwidth,height=0.3\textwidth]{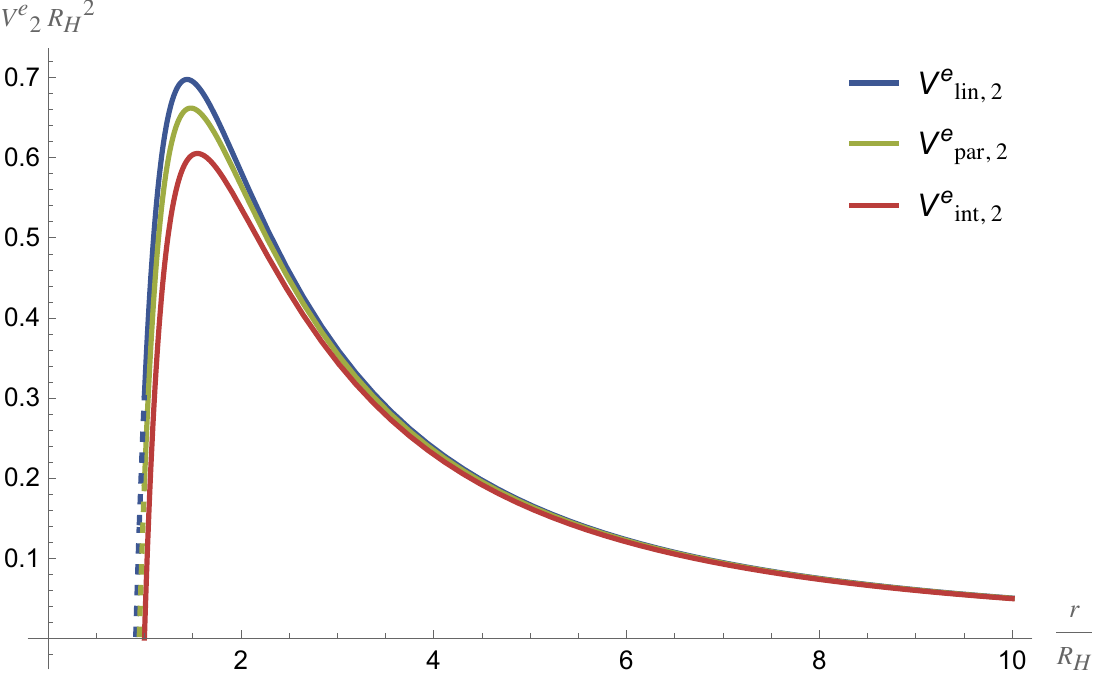}
$\qquad$
\includegraphics[width=0.43\textwidth,height=0.3\textwidth]{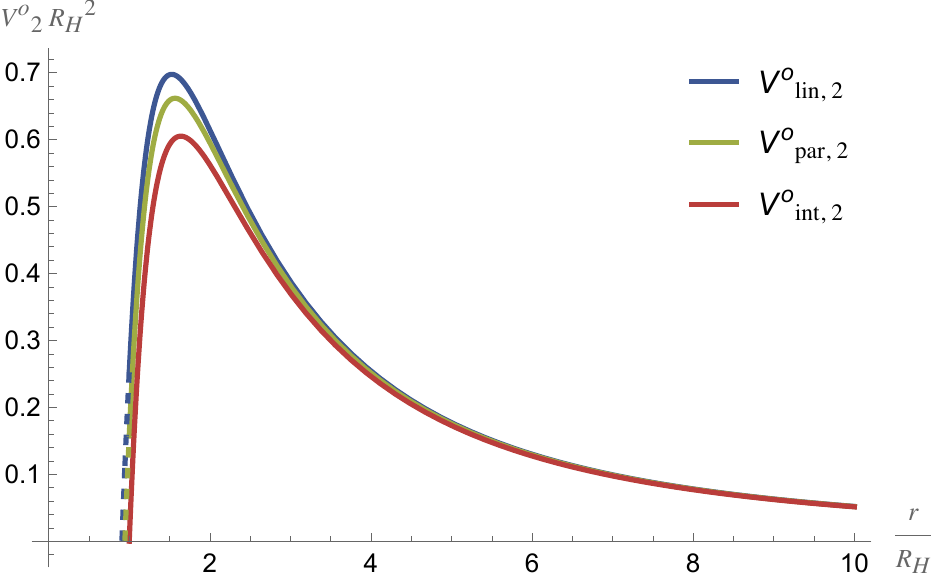}
\caption{Potentials for scalar perturbations $V_0\,\Rh^2$ (top left), vector perturbations $V_1\,\Rh^2$ (top right),
even tensor perturbations $V^{\mathrm{(e)}}_2\, \Rh^2$ (bottom left), and odd tensor perturbations $V^{\mathrm{(o)}}_2\,\Rh^2$
(bottom right): $V_{\rm lin}$ for the linear mass function~\eqref{m_lin},
$V_{\rm par}$ for the parabolic mass function~\eqref{m_par}, and $V_{\rm int}$ for the interpolating
mass function~\eqref{eq:massfunctioninterpolated}.
All plots with $M=500\,\mpl$ and $\mu=1/100\,\mpl$.}
 \label{fig:V2}
\end{figure}

These potentials are shown in Fig.~\ref{fig:V} and Fig.~\ref{fig:V2}, for the effective mass functions $m_{\rm lin}$,
$m_{\rm par}$ and $m_{\rm int}$ of the outermost layer of the dust core introduced in Section~\ref{sec2.2}.
In particular, we recall that the interpolating mass function~\eqref{eq:massfunctioninterpolated} exactly
reproduces the vacuum Schwarzschild geometry where QNMs have support. 
By comparing the plots in Fig.~\ref{fig:V} with those in Fig.~\ref{fig:V2}, it is clear that the differences 
with respect to the Schwarzschild case become smaller for larger values of the ADM mass $M\gg\mu$.
This is expected from the asymptotic behaviour of the wave-functions~\eqref{psi} for $r\gg \bar R_i\sim M_i$.
\par
Having specified the potentials, one can solve the master equation~\eqref{KGuS}
with the proper boundary conditions and obtain the QNMs frequencies. 
Several numerical methods can be employed, based on convergent algorithms,
that require non trivial analysis of the singular points of the master equation. 
Moreover, some methods involve different procedures for different master equations,
which make them less convenient in general. 
Conversely, the WKB method proposed in Ref.~\cite{Iyer:1986vv} is suitable for various
master equations and relative potentials and has been extended to third order in
Refs.~\cite{Iyer:1986np,Iyer:1986nq} and lately improved to sixth and thirteenth
order~\cite{Konoplya:2011qq,Matyjasek:2017psv}. 
We recall that in our model the black hole is spherically symmetric and the potential
is asymptotically constant.
Under these conditions, the main idea is to treat the master equation as a
Schr\"odinger-like equation, with a barrier $Q(r)$ with one peak and two roots,
that divide the real line in three regions: I and III outside the turning points and II between them.
The solutions in these regions have to be carefully connected with a Taylor expansion
near the peak of the potential.
In tortoise coordinates, region~I corresponds to spatial infinity ($r_* \to +\infty$),
whereas region~III corresponds to the horizon $r=\Rh$ ($r_* \to -\infty$).
The crucial part is to connect the amplitudes at infinity with those at the horizon,
which is done starting from the usual WKB asymptotic expansion valid in region I and III:
\be
\xi(r_*) \sim \exp{\sum_{n=0}^{\infty}\frac{\epsilon^n\, S_n(r_*)}{\epsilon}}
\ ,
\label{WKB}
\ee
where the parameter $\epsilon$ keeps track of the expansion order.
Once~\eqref{WKB} is replaced into the master equation$~\eqref{KGuS}$~\footnote{The
second order term in Eq.~$\eqref{KGuS}$ is to be considered multiplied by $\epsilon^2$.}
one obtains the expressions for the $S_n$.
The first two terms $S_0$ and $S_1$ provide a solution of the form
\be
\xi(r_*)
\sim 
\begin{cases}
A^{(\mathrm{in})}_{\infty}\,e^{-i\,k_{\infty}r_{*}}
+
A^{(\mathrm{out})}_{\infty}\,e^{+i\,k_{\infty}r_{*}}
\ ,
\qquad
k^2_{\infty}=
\omega^2
-
\displaystyle
\lim_{r_*\to+\infty} V(r_*)
\\
\\
A^{(\mathrm{in})}_{\rm H}\,e^{+i\,k_{\rm H}r_{*}}
+
A^{(\mathrm{out})}_{\rm H}\,e^{-i\,k_{\rm H}r_{*}}
\ ,
\qquad
k^2_{\rm H}
=
\omega^2
-
\displaystyle\lim_{r_*\to-\infty}V(r_*)
\ ,
\end{cases}
\ee
with $A_{\rm H}$ and $A_{\infty}$ complex amplitudes at the horizon and at infinity,
respectively.
The amplitudes are linearly related according to
\be
\begin{pmatrix}
A^{(\mathrm{out})}_{\rm H}
\\
A^{(\mathrm{in})}_{\rm H}
\end{pmatrix}
=
\begin{pmatrix}
M_{11} & M_{12}
\\
M_{21} & M_{22}
\end{pmatrix}
\begin{pmatrix}
A^{(\mathrm{out})}_{\infty}
\\
A^{(\mathrm{in})}_{\infty}
\end{pmatrix}
\label{matrix}
\ee
and the next goal is to determine the matrix elements $M_{ij}$.
The final step consists into matching WKB solutions in region~I and~III with the one in region~II.
This is done by Taylor expanding $Q(r_*)$ around its maximum and solve equation~\eqref{KGuS}
with a combination of parabolic cylinder functions. 
The coefficients of these asymptotically expanded solutions, can be matched with those
in the outer region and used to populate the matrix~\eqref{matrix}.
The boundary conditions for $\xi$ at $r_* \to \pm \infty$ follow from requiring
that no waves can escape from the horizon, {\em i.e.}~$A^{(\mathrm{in})}_{\rm H}=0$,
and that there are no incoming waves, {\em i.e.}~$A^{(\mathrm{in})}_{\infty}=0$.~\footnote{These
arguments are reversed if $\omega_R<0$.}
These conditions on the amplitudes forces the respective coefficients $M_{ij}$ to vanish,
which finally allows us to derive the expression
\be
i\, \frac{\omega^2-V_s(\bar{r}_*)}{\sqrt{-2\,V''_s(\bar{r}_*)}}
=
n+\frac{1}{2}
\ ,
\qquad
n=0,1,2,\ldots
\ ,
\label{omega0}
\ee
with the integer $n$ called the overtone number and $\bar{r}_*$ denotes the location
of the maximum of $V_s$.
Eq.~\eqref{omega0} provides a way to compute the QNMs frequencies $\omega$. 
\par
In Refs.~\cite{Konoplya:2011qq,Matyjasek:2017psv}, this approach has been extended
to higher orders yielding:
\be
i\, \frac{\omega^2-V_s(\bar{r}_*)}{\sqrt{-2\,V''_s(\bar{r}_*)}}
-
\sum_{i=2}^{k}\Lambda_i
=
n+\frac{1}{2}\ ,
\qquad
n=0,1,2,\ldots
\ ,
\label{omega}
\ee
with $\Lambda_i$ that depends on the effective potential and its derivative and
accounts for higher-order corrections.
Such corrections are needed for achieving greater accuracy.
For this reason, the authors of Ref.~\cite{Konoplya:2011qq} extended the WKB method
to the thirteenth order and introduced the Pad\'e approximants that sensibly improved
the precision for the frequencies, especially for $n\leq \ell$.
On top of that, for a generic order $k$ and a given Pad\'e approximant
$P_{\tilde{n}/ \tilde{m}}$ such that $k=\tilde{n}+\tilde{m}$, the best results are typically
obtained for $\tilde{n} \sim \tilde{m}$. 
However, it is important to mention that higher-order corrections do not always correspond
to a convergence of the method or to more accurate results. 
The degree of accuracy depends fon the errors associated with the computation,
that are usually estimated as the maximum deviation between the average frequency
$\bar{\omega}_k$ at order $k$, obtained from all Pad\'e approximants $P_{\tilde{n}/ \tilde{m}}$
with $\tilde{n}+\tilde{m}=k$, and the individual values $\omega_{\tilde{n}/ \tilde{m}}$. 
This choice tracks the stability of the solutions and how the values of the approximated
frequencies change across different orders, which is reasonable from a convergence perspective,
even if it is not mathematically rigorous.
Nevertheless, this setup generally offers the best compromise between the overall
validity of the method and the quality of the results, especially for higher overtone numbers~\cite{Konoplya:2019hlu}. 
\par
Given the above remarks, we computed the QNMs using the Pad\'e approximants to thirteenth
order for all the models, except for the interpolating case (equivalent to the standard Schwarzschild
vacuum), where we limited the analysis to fifth order.
The numerical values of the QNM frequencies for $M=\mpl=10\,\mu$ (corresponding
to the potentials in Fig.~\ref{fig:V}) are displayed in Tables~\ref{T1}-\ref{T4}.
Both linear and parabolic mass functions lead to small deviations for the frequencies
from the Schwarzschild case, with the parabolic model yielding frequencies closer
to the Schwarzschild ones, compared to the linear case, across all perturbation types.
We recall that this case represents a more accurate internal mass distribution, with a smaller
fraction of ADM mass near the surface of the core.
\par
We also remark that the calculations for the interpolating case are performed to fifth order for
practical considerations, since higher orders of Pad\'e approximants become more computationally
demanding than advantageous.
For some values of $n$ and $\ell$,~\footnote{Specifically when $n\geq\ell$ where the method
is less accurate.} the polynomials involve very small numbers that compromise the numerical
stability of their asymptotic convergence, rather than improving the precision.
This is not a real drawback, as fifth-order frequencies are very accurate and there is no a priori
reason to force higher-order corrections~\cite{Konoplya:2019hlu}.
%
%
%
\begin{table}
\begin{center}
  \def\arraystretch{1.4}
  \begin{tabular}{|c||c|c|c|}
    & Schwarzschild/Interpolating & Linear & Parabolic  \\ \hline \hline
    $n=0,\ \ell=0$ & $0.111 - 0.105\,i$ & $0.148 - 0.138\,i$ & $0.132 - 0.125\,i$\\
    \hline
    $n=0,\ \ell=1$ & $0.293 - 0.0977\,i$ & $0.391 - 0.130\,i$ & $0.349 - 0.116\,i$ \\ 
    \hline
    $n=1,\ \ell=1$ & $0.264 - 0.306\,i$ & $0.352 - 0.409\,i$ & $0.315 - 0.364\,i$ \\
    \hline
    $n=0,\ \ell=2$ & $0.484 - 0.0968\,i$ & $0.645 - 0.129\,i$ & $0.575 - 0.115\,i$\\
    \hline
    $n=1,\ \ell=2$ & $0.464 - 0.296\,i$ & $0.618 - 0.394\,i$ & $0.552 - 0.352\,i$ \\ 
    \hline
    $n=2,\ \ell=2$ & $0.431 - 0.509\,i$ & $0.574 - 0.678\,i$ & $0.512 - 0.605\,i$  \\
    \hline
    $n=0,\ \ell=3$ & $0.675 - 0.0965\,i$ & $0.900 - 0.129\,i$ & $0.804 - 0.115\,i$ \\ 
    \hline
    $n=1,\ \ell=3$ & $0.661 - 0.292\,i$ & $0.881 - 0.390\,i$ & $0.786 - 0.348\,i$ \\
    \hline
    $n=2,\ \ell=3$ & $0.634 - 0.496\,i$ & $0.845 - 0.661\,i$ & $0.754 - 0.590\,i$ \\
    \hline
    $n=3,\ \ell=3$ & $0.599 - 0.711\,i$ & $0.798 - 0.948\,i$ & $0.712 - 0.846\,i$ \\ \hline
  \end{tabular}
  \caption{Scalar perturbations.
  \label{T1}}
\end{center}
\end{table}
\begin{table}
\begin{center}
  \def\arraystretch{1.4}
  \begin{tabular}{|c||c|c|c|}
    & Schwarzschild/Interpolating & Linear & Parabolic  \\ \hline \hline
    $n=0,\ \ell=1$ & $0.248 - 0.0925\,i$ & $0.331 - 0.123\,i$ & $0.295 - 0.110\,i$\\
    \hline
    $n=1,\ \ell=1$ & $0.215 - 0.294\,i$ & $0.285 - 0.392\,i$ & $0.255 - 0.349\,i$ \\ 
    \hline
    $n=0,\ \ell=2$ & $0.458 - 0.0950\,i$ & $0.610 - 0.127\,i$ & $0.544 - 0.113\,i$\\ 
    \hline
    $n=1,\ \ell=2$ & $0.437 - 0.291\,i$ & $0.582 - 0.388\,i$ & $0.519 - 0.346\,i$ \\
    \hline
    $n=2,\ \ell=2$ & $0.401 - 0.502\,i$ & $0.535 - 0.668\,i$ & $0.477 - 0.597\,i$\\ 
    \hline
    $n=0,\ \ell=3$ & $0.657 - 0.0956\,i$ & $0.876 - 0.127\,i$ & $0.782 - 0.113\,i$\\
    \hline
    $n=1,\ \ell=3$ & $0.642 - 0.290\,i$ & $0.856 - 0.386\,i$ & $0.764 - 0.345\,i$ \\ 
    \hline
    $n=2,\ \ell=3$ & $0.614 - 0.492\,i$ & $0.818 - 0.656\,i$ & $0.730 - 0.586\,i$ \\ 
    \hline
    $n=3,\ \ell=3$ & $0.578 - 0.706\,i$ & $0.770 - 0.941\,i$ & $0.688 - 0.840\,i$ \\
    \hline
  \end{tabular}
  \caption{Vector perturbations.
  \label{T2}}
\end{center}
\end{table}
%
\begin{table}
\begin{center}
  \def\arraystretch{1.4}
  \begin{tabular}{|c||c|c|c|}
    & Schwarzschild/Interpolating& Linear & Parabolic \\ \hline \hline
    $n=0,\ \ell=2$ & $0.374 - 0.0890\,i$ & $0.498 - 0.118\,i$ & $0.445 - 0.106\,i$\\ 
    \hline
    $n=1,\ \ell=2$ & $0.347 - 0.274\,i$ & $0.461 - 0.365\,i$ & $0.412 - 0.326\,i$\\
    \hline
    $n=2,\ \ell=2$ & $0.298 - 0.477\,i$ & $0.397 - 0.638\,i$ & $0.355 - 0.568\,i$\\
    \hline
    $n=0,\ \ell=3$ & $0.599 - 0.0927\,i$ & $0.799 - 0.124\,i$ & $0.713 - 0.110\,i$\\ 
    \hline
    $n=1,\ \ell=3$ & $0.583 - 0.281\,i$ & $0.777 - 0.375\,i$ & $0.693 - 0.335\,i$\\
    \hline
    $n=2,\ \ell=3$ & $0.552 - 0.479\,i$ & $0.735 - 0.639\,i$ & $0.656 - 0.570\,i$\\
    \hline
    $n=3,\ \ell=3$ & $0.512 - 0.690\,i$ & $0.682 - 0.920\,i$ & $0.609 - 0.821\,i$\\
    \hline
  \end{tabular}
  \caption{Odd tensor perturbations.
  \label{T3}}
\end{center}
\end{table}
%
\begin{table}
\begin{center}
  \def\arraystretch{1.4}
  \begin{tabular}{|c||c|c|c|}
    & Schwarzschild/ Interpolating & Linear & Parabolic  \\ \hline \hline
    $n=0,\ \ell=2$ & $0.374 - 0.0890\,i$ & $0.498 - 0.119\,i$ & $0.498 - 0.119\,i$ \\ 
    \hline
    $n=1,\ \ell=2$ & $0.347 - 0.274\,i$ & $0.462 - 0.365\,i$ & $0.462 - 0.365\,i$\\
    \hline
    $n=2,\ \ell=2$ & $0.301 - 0.478\,i$ & $0.401 - 0.635\,i$ & $0.401 - 0.635\,i$\\
    \hline
    $n=0,\ \ell=3$ & $0.599 - 0.0927\,i$ & $0.799 - 0.124\,i$ & $0.799 - 0.124\,i$\\
    \hline
    $n=1,\ \ell=3$ & $0.583 - 0.281\,i$ & $0.777 - 0.375\,i$ & $0.777 - 0.375\,i$\\
    \hline
    $n=2,\ \ell=3$ & $0.552 - 0.479\,i$ & $0.736 - 0.639\,i$ & $0.736 - 0.639\,i$\\
    \hline
    $n=3,\ \ell=3$ & $0.512 - 0.690\,i$ & $0.682 - 0.920\,i$ & $0.682 - 0.920\,i$\\
    \hline
  \end{tabular}
  \caption{Even tensor perturbations.
  \label{T4}}
\end{center}
\end{table}
\section{Conclusion and outlooks}
\label{sec4}
\setcounter{equation}{0}
We started by reviewing the quantum dust core model developed in Ref.~\cite{Casadio:2023ymt}
and its effective energy density $\rho$ in Eq.~\eqref{rho}.
This effective density yields a linear MSH mass in the interior ($r<R_{\rm s}\simeq 3\,\Rh/4$)
shown in Eq.~\eqref{mlinear} and the left panel of Fig.~\ref{N=3 & linear fit}.
Linearity makes the internal metric integrable, and provides a framework for investigating
physically more acceptable black hole models~\cite{Casadio:2023iqt}, whose QNMs can be
investigated. 
\par
However, the linear mass function does not account for the full quantum nature of the
dust particles in the collapsed core.
In particular, it discards the overlapping of the eigenfunctions~$\eqref{psi}$ shown in the
right panel of Fig.~\ref{N=3 & linear fit},
whose squared amplitude weighs the location of the particles inside each layer and
yields the parabolic mass distribution~\cite{Gallerani:2025wjc} shown in Eq.~\eqref{Mnew}
and the left panel of Fig.~\ref{mass profiles}.
In this work, we further refined the description of the outermost layer, to account for the quantum leaking
of the dust particles into the region outside the horizon. 
This is the effect that can lead to modifications of (linear) perturbations in the exterior space.
To show that neglecting this quantum leakage reproduces the Schwarzschild phenomenology,
we also considered the mass function~\eqref{eq:massfunctioninterpolated} that interpolates
smoothly between the interior linear MSH mass and the total ADM mass at a value of the areal
coordinate $r=R_{\rm s}<\Rh$.
\par
Finally, we derived the QNMs potential~\eqref{Vs} that accommodates scalar ($s=0$),
vector ($s=1$) and odd tensor ($s=2$) perturbations, while the potential for even tensor
perturbations is given in Eq.~\eqref{Ve}. 
The corresponding QNM frequencies were determined from Eq.~\eqref{omega} in the framework
of the WKB approximation, following the discussion in Ref.~\cite{Antonelli:2025yol}.
The results are listed and compared to the Schwarzschild spectra in Tables~\ref{T1}-\ref{T4}.
As expected, the interpolating mass function yields identical results to the Schwarzschild black hole,
whereas the MSH mass functions obtained from the explicit wave-functions for dust particles
in the outermost layer lead to sizeable, albeit typically very small, deviations.
In particular, deviations obtained from the parabolic interior profile are usually smaller
than those stemming from the linear case.
This could already be predicted from the shape of the potentials shown in Fig.~\ref{fig:V},
since the linear mass function leads to larger deviations from the potentials for the
Schwarzschild geometry.
It also agrees with the fact that the amount of dust in the outermost layer
is larger for the linear profile than for the parabolic case, which reduces the
total mass that can leak outside the horizon.
\subsection*{Data Availability Statement}
Data sharing was not applicable to this article as no data sets were generated or analyzed in this study.
\subsection*{Conflict of Interest}
The authors declare no conflict of interest.
\subsection*{Acknowledgments}
We would like to thank T. Antonelli for helpful discussions. L.G., A.M., A.G. and R.C.~carried out this work in the framework of the activities of the Italian National Group of Mathematical Physics [Gruppo Nazionale per la Fisica Matematica (GNFM), Istituto Nazionale di Alta Matematica (INdAM)].
T.B., A.M., A.G.~and R.C.~are partially supported by the INFN grant FLAG.
R.C.~is also associated with the COST action CA23115 (RQI).
A.M. is partially supported by MUR under the PRIN2022 PNRR project No. P2022P5R22A.
A.G. is supported by the Italian Ministry of Universities and Research (MUR) through the grant ``BACHQ: Black Holes and The Quantum'' (grant no. J33C24003220006).

\appendix
\section{Hermite interpolation}
\label{appendix-a}
\setcounter{equation}{0}
We determine a smooth interpolating mass function of the form in Eq.~\eqref{eq:massfunctioninterpolated}
requiring that $B(r)$, $B'(r)$ and $B''(r)$ be continuous at the boundaries $r=r_0$ and $r=r_1$ to ensure
the continuity of all components of the (effective) energy-momentum tensor at both ends.
This results in the six conditions
\begin{align}
\alpha \, r_0
&=
B(r_0)
\ ,
&
M
&=
B(r_1)
\ ,
\label{eq:cont}
\\ 
\alpha
&=
B'(r_0)
\ ,
&
0
&=
B'(r_1)
\ ,
\label{eq:dercont}
\\ 
0
&
=B''(r_0)
\ ,
&
0
&=
B''(r_1)
\ ,
\label{eq:secdercont}
\end{align}
which can hold for an ``osculating polynomial'' of order $K=5$.
\par
Osculating polynomials are interpolating functions which pass through a set of points with specified
derivatives~\cite{Burden2015,Spitzbart1960}.
Given a set of $n+1$ points $\{x_i\}_{i=0}^{n}$, non-negative numbers $\{ m_i {\}}_{i=0}^{n}$, with
$ m_i \in \mathbb{N} \cup \{ 0 \}$, and values to interpolate $\{ f_{j}^{(k)} \}$, with $j=0, \ldots,n$ and $k=0,\ldots,m_j$,
the osculating polynomial approximating the function $f \in C^{m}( [a,b])$ is the polynomial of least degree such that
\be
\label{eq:conditionsInterpolation}
f^{(k)}(x_j)
=
f_j^{(k)}
\ ,
\qquad
\forall j=0,\ldots,n
\ ,
\quad
\forall k=0,\ldots,m_j
\ .
\ee
From these requirements we have $n+1$ conditions for $f^{(0)}\equiv f$, and $\sum_{j=0}^{n}m_j$
more conditions to be satisfied for the derivatives.
A polynomial of degree
\be
\label{eq:polorder}
K
=
\sum_{j=0}^{N}m_j + n
\ee
has $K+1$ coefficients that can be used to fulfil these requirements.
A general theorem~\cite{Burden2015} states that a polynomial of degree $K$ such that
the conditions~\eqref{eq:conditionsInterpolation} are satisfied exist and is given by
\be
\label{eq:interpolatingPolynomial}
f(x)
=
\sum_{j=0}^{n} \sum_{k=0}^{m_j} A_{jk}(x)\, f_j^{(k)}
\ ,
\ee
where
\be
A_{jk}(x)
=
p_j(x)\,\frac{\left(x-x_j\right)^k}{k!}
\,\sum_{l=0}^{m_j-k}
\frac{g_j^{(l)}(x_j)}{l!}\left(x-x_j\right)^l
\ ,
\ee
with 
\begin{align}
 p_j(x)
 &
 =
 \prod_{ \substack{l=0, l \neq j}}^{n} 
 \left(x-x_l\right)^{m_l+1}
 \\
g_j(x)
&=
\frac{1}{p_j(x)}
\ .
\end{align}
In the special case $m_j=1$ for each $j=0,\ldots,n$, the polynomials $f(x)$ are usually called Hermite
polynomials and Eq.~\eqref{eq:interpolatingPolynomial} is called the Hermite formula.
\par
In our case, we have conditions up to second derivatives at the two points $x_0$ and $x_1$,
so that $m_0=m_1=2$.
The general formula~\eqref{eq:interpolatingPolynomial} hence simplifies to
\be
B(x)
=
\sum_{j=0}^{1} \sum_{k=0}^{2} A_{jk}(x) \,B_j^{(k)}
\ ,
\ee
where $x=r/R_H$, $x_0=r_0/R_H$ and $x_1=r_1/R_H$ and 
\be
B_j=B^{(0)}(x_j) \quad B'_j=B^{(1)}(x_j) \quad B''_j=B^{(2)}(x_j)
\ .
\ee
\par
The conditions~\eqref{eq:cont}, \eqref{eq:dercont} and \eqref{eq:secdercont} imply
\be
B^{(1)}(x_1)=B^{(2)}(x_0)=B^{(2)}(x_1)=0
\ ,
\ee
and we can set
\be
\label{eq:functionspol}
A_{02}(x)=A_{11}(x)=A_{12}(x)=0
\ .
\ee
The full problem therefore simplifies to determine only the three functions $A_{00}(x)$, $A_{01}(x)$ and $A_{10}(x)$.
Assuming $r_0 = \bar{R}_{N}$ and $ r_1=\bar{R}_{N}+\Delta R_{N}$, we define
\be
\Delta x = \frac{\Delta R_{N}}{\Rh}
\  ,
\ee
and find that these functions read
\begin{align}
A_{00}(x)
=
&
- \frac{1}{\Delta x^3} (x-x_1)^3 - \frac{3}{ \Delta x^4} (x-x_1)^3 (x-x_0)
- \frac{6}{\Delta x^5}(x-x_1)^3(x-x_0)^2
\\
A_{01}(x)
=
&
- \frac{1}{\Delta x^3}(x-x_1)^3(x-x_0)-\frac{3}{\Delta x^4}(x-x_1)^3(x-x_0)^2
\\
A_{10}(x)
=
&
+ \frac{1}{\Delta x^3}(x-x_0)^3 - \frac{3}{\Delta x^4}(x-x_0)^3(x-x_1)
+\frac{6}{\Delta x^5}(x-x_1)^2(x-x_0)^3
\ .   
\end{align}
Finally, the complete interpolating function is given by
\begin{align}
B(x)
=
&
-\frac{c_1}{\Delta x^3}(x-x_1)^3+\frac{M}{\Delta x^3}(x-x_0)^3 
-\left( \frac{3 c_1}{\Delta x^4} + \frac{c_2}{\Delta x^3} \right)
(x-x_1)^3(x-x_0)
\notag
\\
&
- \frac{3M}{\Delta x^4}(x-x_0)^3(x-x_1)
-\left( \frac{6c_1}{\Delta x^5} + \frac{3c_2}{\Delta x^4} \right)
(x-x_1)^3(x-x_0)^2
\notag
\\
&
+ \frac{6M}{\Delta x^5}(x-x_1)^2(x-x_0)^3
\ ,
\end{align}
with $c_1= M_N$ and $c_2= c_1/x_0$.
%
%

\end{document}